# Reply to "Comments on Integer SEC-DED codes for low power communications"

Aleksandar Radonjic and Vladimir Vujicic, *Member, IEEE*

*Abstract*—This paper is a reply to the comments on "Integer SEC-DED codes for low power communications".

## I. Introduction

A reader of the paper [2], who is not familiar with error control coding, may conclude that some results from [1] are incorrect. In this paper we show that these results are correct, i.e. that the statements given in [2] are misleading due to the use of wrong and/or non-existent expressions.

## II. On the code complexity

Commenting on complexity of encoding and decoding procedures of integer SEC-DED codes, authors of [2, Section 2] state that from the relation $k=2^{b-2}$ ($k$ – number of databytes, $b$ – databyte length) it follows that $b = ld(k)+2$, i.e. that time complexity of encoding and decoding procedures is not $O((4b+1) \cdot k) \approx 4b \cdot O(k)$, but $O(k \cdot (4ld(k)+1)) \approx O(k \cdot ld(k))$.

This conclusion is the result of incorrect interpretation of the meaning of the parameter $k$, whose value has no impact on the time complexity of encoding and decoding procedures. Namely, in [1, Section 4] it has been shown that the integer SEC-DED encoder/decoder must perform $4b+1$ operations at byte level, i.e. $(4b+1) \cdot k$ operations in case of $k$ bytes. From this it is obvious that the total number of operations linearly depends on $k$, regardless of whether it is $k \leq 2^{b-2}$ or $k > 2^{b-2}$. However, if we want to construct integer codes with SEC-DED capabilities, we have to take into account the limitation regarding the number of databytes ($k \leq 2^{b-2}$).

Another incorrect statement of the paper [2] concerns the complexity of encoding and decoding procedures of linear SEC-DED codes [2, Section 2]. Namely, the authors of [2] claim that in [3, Theorems 2.4 and 2.5] it has been proved that the sequential encoding/decoding complexity of generalized SEC-DED codes is $k \cdot ld(k)$, which is actually not true. The reader can easily be convinced that [3, Theorems 2.4 and 2.5] is concerned with the complexity of iterated majority voting algorithm (Theorem 2.4) and parallel iterated majority voting algorithm (Theorem 2.5), and not with the complexity of generalized SEC-DED codes.

## III. On the expression for the first check-byte

In the last paragraph of their paper the authors of [2] state that our expression for the first check-byte from [1] is the same as those in [4], [5] which is also not true. This expression, also known as the arithmetic checksum, is firstly defined by Wakerly in his paper of 1976 [6], [7, p. 560].

The authors are with the Faculty of Technical Sciences, University of Novi Sad, Serbia (e-mail: sasa_radonjic@yahoo.com, vujicicv@uns.ac.rs).